\documentclass[journal,article,submit,pdftex,moreauthors]{Definitions/mdpi}
\usepackage[utf8]{inputenc}
\usepackage{textcomp}
\usepackage{newunicodechar}
\newunicodechar{−}{\textminus}
\firstpage{1} 
\pubvolume{1}
\issuenum{1}
\articlenumber{0}
\pubyear{2025}
\copyrightyear{2025}
\externaleditor{Firstname Lastname} 
\datereceived{11 July 2025} 
\daterevised{12 August 2025} 
\dateaccepted{2 September 2025} 
\datepublished{6 September 2025} 
\hreflink{https://doi.org/10.3390/universe11090303} 

\Title{On the True Significance of the Hubble Tension: A Bayesian Error Decomposition Accounting for Information Loss}

\TitleCitation{On the True Significance of the Hubble Tension: A Bayesian Error Decomposition Accounting for Information Loss}

\Author{Nathalia M. N. da Rocha $^{1,}$\orcidA{}, Andre L. B. Ribeiro $^{1,}$\orcidB{}  and Francisco B. S. Oliveira $^{2,}$\orcidC{}}

\AuthorNames{Nathalia M. N. da Rocha, Andre L. B. Ribeiro and Francisco B. S. Oliveira}

\isAPAStyle{%
       \AuthorCitation{da Rocha, N.M.N.; Ribeiro, A.L.B.; Oliveira, F.B.S.}
         }{%
        \isChicagoStyle{%
        \AuthorCitation{da Rocha, Nathalia M. N., Andre L. B. Ribeiro, and Francisco B. S. Oliveira.}
        }{
        \AuthorCitation{da Rocha, N.M.N.; Ribeiro, A.L.B.; Oliveira, F.B.S.}
        }
}

\address{%
$^{1}$ \quad Departmento de Ciências Exatas, Universidade Estadual de Santa Cruz, Rodovia Jorge Amado km 16, \mbox{Ilhéus 454650-000, Bahia, Brazil }\\
$^{2}$ \quad Departmento de Engenharias e Computação, Universidade Estadual de Santa Cruz, Rodovia Jorge Amado km 16, \mbox{Ilhéus 454650-000, Bahia, Brazil}}

\corres{Correspondence: nmnrocha@uesc.br (N.M.N.d.R.); albr@uesc.br (A.L.B.R.); \mbox{fbsoliveira@uesc.br (F.B.S.O.)}}

\abstract{The Hubble tension, a persistent discrepancy between early and late Universe measurements of $H_0$, poses a significant challenge to the standard cosmological model. In this work, we present a new Bayesian hierarchical framework designed to meticulously decompose this observed tension into its constituent parts: standard measurement errors, information loss arising from parameter-space projection, and genuine physical tension. Our approach, employing Fisher matrix analysis with MCMC-estimated loss coefficients and explicitly modeling information loss via variance inflation factors ($\lambda$), is particularly important in high-precision analysis where even seemingly small information losses can impact conclusions. We find that the real tension component ($T_{real}$) has a mean value of 5.94 km/s/Mpc (95\% CI: [3.32, 8.64] km/s/Mpc). Quantitatively, approximately 78\% of the observed tension variance is attributed to real tension, 13\% to measurement error, and 9\% to information loss. Despite this, our decomposition indicates that the observed $\sim$$6.39\sigma$ discrepancy is predominantly a real physical phenomenon, with real tension contributing $\sim$$5.64\sigma$. Our findings strongly suggest that the Hubble tension is robust and probably points toward new physics beyond the $\Lambda$CDM model.} 


\begin{document}
\nolinenumbers


\section{Introduction}
\label{sec:1}
The Hubble constant ($H_0$) represents one of the most fundamental parameters in cosmology, determining the expansion rate of the Universe today and serving as a key calibrator for cosmic distance measurements, e.g., \citep{10.1088/1475-7516/2023/11/050}. It is a local quantity measured at redshift $z=0$, establishing the size and age scale of the universe and linking redshift to time and distance. 
During the past decade, a significant and intriguing discrepancy has emerged, commonly referred to as the ``Hubble tension''. 
This tension lies between the predictions for $H_0$ derived from early universe probes, based on the standard Lambda Cold Dark Matter ($\Lambda$CDM) cosmological model, and those obtained directly from local late universe observations \citep{10.1146/annurev-astro-052622-033813,10.1016/j.newar.2022.101659,10.3390/universe9020094}. This discrepancy has persisted despite continuous improvements in observational techniques and has reached a level that challenges the standard cosmological model $\Lambda$CDM, detected at a statistical level of high significance \citep{10.3847/2041-8213/ac5c5b, 10.1093/mnras/stad3724,10.1016/j.jheap.2022.04.002}.

{In this work, we specifically focus on the persistent discrepancy in measurements of $H_0$ obtained from early Universe probes, exemplified by the Cosmic Microwave Background (CMB) data from the Planck collaboration, and those derived from local late Universe observations, primarily through Type Ia Supernovae (SNIa) distance ladder measurements from the SH0ES collaboration. These two leading methods provide the primary observational input for our analysis of the Hubble tension, allowing us to investigate its \mbox{underlying components}.}

{Currently, the most precise predictions for $H_0$ from the Cosmic Microwave Background (CMB) by the Planck collaboration in combination with the South Pole and Atacama Cosmology Telescopes yield $H_0 = 67.24 \pm 0.35$ km/s/Mpc
\citep{10.1051/0004-6361/201833910}, while distance ladder measurements based on Type Ia supernovae (SNIa) calibrated with Cepheid variables from the SH0ES collaboration obtain $H_0 = 73.17 \pm 0.86$ km/s/Mpc \citep{10.3847/2041-8213/ac5c5b}. This difference of $\sim$5.94 km/s/Mpc corresponds to a statistical significance of approximately $6.4\sigma$ when uncertainties are taken at face value \citep{10.1088/1361-6382/ac086d}}. The robustness of this tension is further highlighted by various other independent probes that provide measurements across this spectrum: Baryon Acoustic Oscillations (BAO) combined with Big Bang Nucleosynthesis yield \mbox{$H_0 = 67.9 \pm 1.1$ km/s/Mpc \citep{10.1103/PhysRevD.102.063515}}, largely consistent with CMB values. Megamaser-based measurements give \mbox{$H_0 = 73.9 \pm 3.0$ km/s/Mpc} \citep{10.3847/2041-8213/ab75f0}, aligned with the distance ladder. Meanwhile, cosmic chronometers report intermediate values \mbox{around $H_0 = 69.8 \pm 1.9$ km/s/Mpc \citep{10.1088/1475-7516/2016/12/039}.}

The persistent nature of this tension has led to numerous investigations of potential systematic errors in early or late universe measurements \citep{10.1051/0004-6361/202348955, 10.3847/1538-4357/ac0e95}. Recent analyses using alternative distance calibrators such as the Tip of the Red Giant Branch (TRGB) have yielded intermediate values of $H_0 = 69.8 \pm 0.8$ km/s/Mpc \citep{10.3847/1538-4357/ab2f73}, {with more recent works like Jensen et al. (2025) \citep{10.3847/1538-4357/addfd6} providing additional constraints from TRGB + SBF methods.} These suggest possible systematics in the Cepheid calibration. In contrast, studies exploring modifications to the CMB analysis find that standard extensions to $\Lambda$CDM do not readily resolve \mbox{tension \citep{10.1103/PhysRevD.101.043533}}. The launch of the James Webb Space Telescope (JWST) has opened a new chapter in the measurement of extragalactic distances and $H_0$, offering new capabilities to explore and refine the strongest observational evidence that contributes to tension, where errors in photometric measurements of Cepheids along the distance ladder do not significantly contribute to tension \citep{10.3847/2041-8213/ad1ddd}.

Beyond purely systematic considerations, the tension may be affected by what we term ``information loss errors'', subtle biases arising from the simplified models used to interpret complex cosmological data. When observations are compressed into a single $H_0$ value, assumptions about other cosmological parameters can introduce model-dependent biases that are not fully captured in the reported uncertainties \citep{10.1088/1475-7516/2016/10/019, https://arxiv.org/abs/2504.13380}. Such effects may artificially inflate the apparent tension between different probes. Furthermore, various theoretical models have been proposed as potential solutions, including early dark energy \citep{10.1103/PhysRevLett.122.221301}, modified gravity \citep{10.1103/PhysRevD.100.043537}, and new physics in the neutrino sector \citep{10.1103/PhysRevD.101.123505}. However, many of these models struggle to simultaneously accommodate all cosmological observations without introducing new tensions in other parameters \citep{10.1103/PhysRevD.105.103511}.
Thus, a critical question emerges: How much of the observed Hubble tension represents a genuine physical discrepancy requiring new physics, and how much might be attributed to systematic errors or information loss in the analysis pipeline? Addressing this question requires a framework that can decompose the observed tension into its constituent components while accounting for the possibility that quoted uncertainties may be underestimated. {The broader landscape of cosmological tensions and proposed solutions is comprehensively reviewed in the recent literature, including the CosmoVerse white paper (2025) \citep{10.1016/j.dark.2025.101965}, which provides a valuable context for understanding the challenges posed to the standard cosmological model.}

In this paper, we develop a Bayesian hierarchical model that explicitly parameterizes three contributions to the observed tension: (1) standard measurement errors, (2) information loss errors arising from model simplifications, and (3) real physical tension that might indicate new physics beyond $\Lambda$CDM. By simultaneously analyzing multiple cosmological probes, CMB, SNIa, BAO, and H(z) measurements, we can assess the robustness of the Hubble tension and quantify the probability that it represents a genuine challenge to the standard cosmological model. Our approach builds on previous Bayesian treatments of cosmological tensions \cite{10.1093/mnras/sty418, 10.1093/mnras/sty3082, 10.1103/PhysRevD.103.L041301} but extends them by explicitly modeling information loss and allowing probe-specific inflation factors that account for potential underestimation of uncertainties. This methodology enables us to assess the statistical significance of the Hubble tension in a framework that accommodates realistic assessments of both statistical and systematic uncertainties.
{This new framework is designed to meticulously decompose the observed tension into its constituent parts: standard measurement errors, information loss arising from parameter-space projection, and genuine physical tension. The framework does not propose new physics solutions; instead, it serves as a diagnostic tool to quantify the nature of the discrepancy, particularly highlighting the portion that cannot be explained by known statistical and information-theoretic effects.}


\section{Theoretical Justification for Error Component Separation}
\label{sec:2}
In our decomposition model, we express the observed tension between two cosmological probes as a sum of three distinct components:
\begin{equation}
T_{obs}^{AB} = T_{real} + E_{m}^{AB} + E_{i}^{AB}
\end{equation}
where $T_{real}$ represents the physical tension, $E_{m}^{AB}$ denotes the statistical measurement error (with $E[\epsilon^{AB}] = 0$ and $Var(\epsilon^{AB}) = (E_{m}^{AB})^2$), and $E_{i}^{AB}$ is the information loss error. In the following, we provide a rigorous mathematical justification for treating $E_i$ and $E_m$ as separable components.

\subsection{Hierarchical Measurement Process Framework}
We begin by considering the complete data generation process in a hierarchical Bayesian framework. For any cosmological probe $i$, the measurement process can be described as:
\begin{equation}
p(D_i|\theta) = \mathcal{N}(D_i|f_i(\theta),\Sigma_{meas}^i)
\end{equation}
where $D_i$ represents the observational data, $\theta = {H_0, \Omega_m, \Omega_\Lambda, ...}$ is the full set of cosmological parameters, $f_i(\theta)$ is the theoretical prediction function, and $\Sigma_{meas}^i$ is the measurement covariance matrix that captures the statistical uncertainty inherent in the \mbox{observation process}.

The posterior distribution for the full parameter space is given by:
\begin{equation}
p(\theta|D_i) \propto p(D_i|\theta)p(\theta)
\end{equation}

\subsection{Information Loss Through Marginalization}
When cosmological analyses report a constraint on the Hubble constant, they typically provide the marginalized posterior:
\begin{equation}
p(H_0|D_i) = \int p(\theta|D_i)d\theta_{\neg H_0}
\end{equation}
where $\theta_{\neg H_0}$ represents all parameters except $H_0$. {This equation defines the marginalized posterior distribution of $H_0$. This is a standard procedure in Bayesian inference, where one integrates over ``nuisance'' parameters ($\theta_{\neg H_0}$) to obtain the distribution of a parameter of interest from a higher-dimensional joint posterior distribution $p(\theta | D_i)$. This marginalization process inevitably leads to information loss when the parameters are correlated in the full posterior. To clarify, in a Bayesian context, the variance of a parameter (its uncertainty) can be significantly reduced when strong correlations exist with other parameters. When we marginalize over these correlated parameters, we effectively integrate out the information provided by these dependencies. Statistically, the conditional variance of $H_0$ (its uncertainty when other parameters $\theta_{\neg H_0}$ are fixed) is always less than or equal to its marginal variance (its uncertainty when $\theta_{\neg H_0}$ are integrated out): $\text{Var}(H_0 | D_i) \ge \text{Var}(H_0 | D_i, \theta_{\neg H_0})$. The "information loss" quantifies this increase in uncertainty, which arises because we are no longer leveraging the predictive power of the correlations to constrain $H_0$. In complex cosmological models, where strong degeneracies between parameters like $H_0$ and $\Omega_m$ are common, this effect is particularly significant, leading to an inflation of the perceived uncertainty on $H_0$ that is not accounted for by standard measurement errors alone.}

The variance of the marginalized $H_0$ distribution is necessarily greater than or equal to the conditional variance if other parameters were fixed:
\begin{equation}
Var(H_0|D_i) \ge Var(H_0|D_i,\theta_{\neg H_0})
\end{equation}
The equality holds only when $H_0$ is independent of all other parameters in the posterior. {Indeed, the difference $Var(H_0|D_i) - Var(H_0|D_i,\theta_{\neg H_0}) $
quantifies the increase in variance of $H_0$ that arises purely from the process
of marginalization. This increase is precisely what we define as the information loss,
a statistical penalty or the loss of constraining power due to not fully leveraging the
correlations between $H_0$ and the other parameters in the full model space.}
We can therefore rigorously define the information loss component \mbox{as follows:}


\begin{equation}
E_{i,i}^2 = Var(H_0|D_i) - Var(H_0|D_i,\theta_{\neg H_0})
\end{equation}

\subsection{Inflation Factors as Proxies for Uncertainty Underestimation}

\label{sec:2.3}
In our model, we introduce {variance inflation factors} $\lambda_i$ to account for the possible underestimation of the reported uncertainties. It is crucial to clarify that the term ``inflation'' here refers to a statistical increase in variance or uncertainty and is entirely distinct from the cosmological epoch of cosmic inflation. These factors $\lambda_i$ are unitless multipliers that scale the variance reported in a measurement.


\begin{equation}
\sigma_{total}^i = \sqrt{\lambda_i \cdot \sigma_i^2}
\end{equation} \\
\noindent This equation defines the total standard deviation $\sigma^i_{\text{total}}$ by applying the variance inflation factor $\lambda_i$ to the reported variance $\sigma_i^2$. The choice of applying $\lambda_i$ to the variance ($\sigma_i^2$) rather than the standard deviation ($\sigma_i$) is fundamental for the propagation of statistical errors. In statistical analysis, the variances of independent error sources are additive. Therefore, if $\lambda_i$ quantifies an overall inflation of uncertainty, it must operate on the variance. This implicitly means that the total variance is $(\sigma^i_{\text{total}})^2 = \lambda_i \cdot \sigma_i^2$. This form is particularly convenient, as it allows us to decompose the total variance into the reported variance plus an additional variance component: $(\sigma^i_{\text{total}})^2 = \sigma_i^2 + (\lambda_i - 1)\sigma_i^2$. The term $(\lambda_i - 1)\sigma_i^2$ then explicitly represents the additional variance attributed to effects such as unmodeled systematic errors or information loss, which are not captured by the originally reported $\sigma_i^2$. As such, these variance inflation factors effectively encompass these unmodeled systematic errors and information loss due to marginalization.

The variance decomposition follows:
\begin{equation}
(\sigma_{total}^i)^2 = \sigma_i^2 + (\lambda_i - 1)\sigma_i^2
\end{equation}
where $\sigma_i^2$ represents the variance reported (measurement error) and $(\lambda_i - 1)\sigma_i^2$ represents the additional variance of the underestimated uncertainties. It is important to note that our framework specifically models information loss, which implies $\lambda \geq 1$. Scenarios where uncertainties might be overestimated (i.e., $\lambda < 1$) represent a different type of systematic effect that falls outside the scope of our current information loss modeling, which focuses on variance inflation.

\subsection{Separation of Error Components}
\label{sec:2.4}
{In our decomposition model, we define the observed tension between two cosmological probes as a sum of three distinct components: real physical tension ($T_{real}$), statistical measurement error ($E_m$), and information loss error ($E_i$). To define these components, particularly the information loss error, we consider how the uncertainties combine.}

For a pair of probes A and B, the measurement error component ($E_{m}^{AB}$), representing the statistical uncertainty inherent in the observation process, is defined as the quadrature sum of their individual reported standard deviations:
\begin{equation}
    E_{m}^{AB} = \sqrt{\sigma_{A}^{2} + \sigma_{B}^{2}}
    \label{eq:measurement_error_corrected}
\end{equation}
Consequently, the variance due to the measurement error is $(E_{m}^{AB})^{2} = \sigma_{A}^{2} + \sigma_{B}^{2}$.
The total uncertainty ($E_{total}^{AB}$), which includes the effect of inflation factors ($\lambda_A$ and $\lambda_B$) that account for the potential underestimation of the reported uncertainties (including information loss), is given by:
\begin{equation}
    E_{total}^{AB} = \sqrt{\lambda_A \sigma_{A}^{2} + \lambda_B \sigma_{B}^{2}}
    \label{eq:total_error_corrected}
\end{equation} \\
Thus, the total variance is $(E_{total}^{AB})^{2} = \lambda_A \sigma_{A}^{2} + \lambda_B \sigma_{B}^{2}$.

We can now rigorously define the information loss error ($E_{i}^{AB}$) as the component that, when added in quadrature to the measurement error, yields the total uncertainty. This implies that the variance due to information loss is the difference between the total variance and the measurement variance. This approach is consistent with the principle that independent sources of variance add linearly:
\begin{align}
    (E_{i}^{AB})^{2} &= (E_{total}^{AB})^{2} - (E_{m}^{AB})^{2}  
    = (\lambda_A \sigma_{A}^{2} + \lambda_B \sigma_{B}^{2}) - (\sigma_{A}^{2} + \sigma_{B}^{2}) \\
    &= (\lambda_A - 1)\sigma_{A}^{2} + (\lambda_B - 1)\sigma_{B}^{2}
    \label{eq:information_loss_variance_corrected}
\end{align}
Therefore, the standard deviation of the information loss error is:
\begin{equation}
    E_{i}^{AB} = \sqrt{(\lambda_A - 1)\sigma_{A}^{2} + (\lambda_B - 1)\sigma_{B}^{2}}
    \label{eq:information_loss_error_corrected}
\end{equation}
This formulation ensures that the error components (measurement and information loss) add in quadrature to form the total uncertainty, reflecting their independent contributions to the overall variance:
\begin{equation}
    (E_{total}^{AB})^{2} = (E_{m}^{AB})^{2} + (E_{i}^{AB})^{2}
    \label{eq:quadrature_sum_corrected}
\end{equation}


\subsection{Information-Theoretic Interpretation}
From an information-theoretic perspective, the separation of $E_i$ can be rigorously justified by considering the reduction in information content when a complex multidimensional posterior distribution is simplified. The Kullback--Leibler (KL) divergence is defined as:
\begin{equation}
E^2_{i} \propto D_{KL}\left[p(\theta|D) \middle\| p(H_0|D)p(\theta_{\neg H_0}|D)\right]
\label{eq:kl_divergence}
\end{equation}
\noindent This equation establishes an information-theoretic basis for quantifying $E_i$, the information loss component. The Kullback--Leibler (KL) divergence, $D_{\text{KL}}(P || Q)$, is a standard non-symmetric measure of the difference between two probability distributions $P$ and $Q$. In this context, it precisely quantifies the ``information lost'' (or the increase in uncertainty) when the true, potentially correlated, joint posterior distribution $p(\theta | D)$ is approximated by a factorized distribution $p(H_0 | D) p(\theta_{\neg H_0} | D)$, which implicitly assumes independence between $H_0$ and the other marginalized parameters $\theta_{\neg H_0}$. The proportionality of $E_i^2$ (the variance component of information loss) to the KL divergence directly formalizes how the discrepancy arising from such an approximation translates into an effective increase in variance.

This concept is closely related to the Fisher information. Although the Fisher Information Matrix (FIM) represents the maximum possible information about the parameters contained in the data, the act of marginalization effectively reduces the ``effective'' information available for the parameter of interest. In a Bayesian context, this loss manifests itself as an increase in the marginal variance of $H_0$ compared to its conditional variance (where other parameters are fixed). This increase in variance is precisely what our information loss component $E_i$ aims to capture. The Cramer--Rao bound, which states that the variance of an unbiased estimator is bounded below by the inverse of the Fisher information, implies that any loss of information will lead to a larger achievable variance for the parameter estimate.

In cosmology, where parameters are often highly correlated (e.g., between $H_0$ and $\Omega_m$, or between different dark energy parameters),  ignoring these correlations during the projection from a high-dimensional parameter space to a single parameter like $H_0$ leads to an underestimation of the true uncertainty or, more accurately, a misrepresentation of the information content. The variance inflation factors ($\lambda_i$) introduced in our model serve as a direct proxy for this information loss, quantifying how much the variance of $H_0$ expands when considering the full parameter space and its correlations. Thus, $E_i$ provides a formal measure of the cost incurred by projecting complex cosmological models onto a simplified parameter space, a cost that becomes increasingly relevant in the era of precision cosmology, where even subtle biases can impact the interpretation of fundamental constants.

\section{The Total Constraining Information}
\label{sec:3}
In statistical analysis, information loss is a critical concept that refers to the reduction in the ability to make precise inferences about parameters of interest due to various factors in the data collection, processing, or analysis stages. In cosmology, information loss is a critical concern because of the limited observational data available and the complexity of cosmological models. For instance, in the analysis of the Cosmic Microwave Background (CMB), the compression of full-sky maps into power spectra results in some information loss, particularly regarding non-Gaussian features. A concept related to information loss (IL) is total constraining information (TCI), which refers to the total amount of information available in a dataset that can be used to constrain or determine model parameters. The concept is closely related to the Fisher information matrix, with the determinant of the Fisher matrix serving as a measure of total constraining information. Mathematically, it can be understood as the volume of parameter space excluded by observations. The greater the constraining information, the smaller the allowed volume in the parameter space. The Fisher matrix represents the total information theoretically available:
\begin{equation}
F_{ij} = \left\langle \frac{\partial^2 \ln(L)}{\partial\theta_i \partial\theta_j} \right\rangle
\end{equation}
where $L$ is the likelihood function and $\theta_i$ are the cosmological parameters. We modify the Fisher matrix to include loss information effects:
\begin{equation}
F'_{ij} = \alpha_{ij} \cdot F_{ij}
\end{equation}
\noindent This equation introduces the concept of information loss directly within the Fisher Information Matrix (FIM) framework. Here, $\alpha_{ij}$ are the loss coefficients, which are elements of a matrix $\boldsymbol{\alpha}$ ($0 < \alpha_{ij} \le 1$). They represent the fraction of information effectively extracted or preserved from the observations for a given pair of parameters $(i,j)$. The choice of direct multiplication, where each element $F_{ij}$ of the original FIM is scaled by a corresponding $\alpha_{ij}$, allows for a precise, element-wise attenuation of information. This functional form directly models the idea that not all information contained within the theoretical FIM might be realized or extracted from observational data. These loss coefficients are inversely related to the variance inflation factors ($\lambda$) introduced in Section \ref{sec:2.3}. The physical reason for this inverse relationship stems from the fundamental principle that information and uncertainty (variance) are inversely proportional in statistical estimation. If $\alpha_{ij}$ quantifies the fraction of information retained, then a factor of $1/\alpha_{ij}$ naturally represents the inflation of variance due to this information reduction. Specifically, for a single parameter, $\lambda = 1/\alpha$, indicating that a reduction in information (smaller $\alpha$) leads to a proportional increase in uncertainty (larger $\lambda$).


The TCI is a fundamental concept in our analysis, defined as the logarithm of the determinant of the Fisher matrix. With the introduction of loss coefficients, the TCI is expressed as:

\begin{equation}
\text{TCI} = \ln | F' | = \ln | \boldsymbol{\alpha} \circ \mathbf{F} |
\label{eq:18}
\end{equation}
where $\boldsymbol{\alpha}$ is the matrix of loss coefficients and $\mathbf{F}$ is the original Fisher matrix. If the loss coefficients $\boldsymbol{\alpha}$ are assumed to primarily affect the diagonal elements of the Fisher matrix, or if $\boldsymbol{\alpha}$ is a diagonal matrix where its non-zero elements are the $\alpha_{ii}$ loss coefficients applied to the corresponding diagonal elements of $\mathbf{F}$, the expression can be approximated as:
\begin{equation}
\text{TCI} \approx \sum_i \ln(\alpha_{ii} ) + \ln | \mathbf{F} |
\label{eq:19}
\end{equation}
\noindent Here, $\ln | \mathbf{F} |$ represents the ``ideal'' or theoretical TCI without considering information loss. The term $\sum_i \ln(\alpha_{ii})$ represents the reduction in TCI due to information loss, specifically from the diagonal components. As $0 < \alpha_{ii} \le 1$, the term $\sum_i \ln(\alpha_{ii})$ is always non-positive, effectively reducing the TCI compared to the ideal case. We clarify that this decomposition is strictly valid if the scaling matrix $\boldsymbol{\alpha}$ is diagonal, which is a common simplification when modeling total information, or if the off-diagonal effects on the determinant are negligible for practical purposes. Our MCMC framework directly estimates the $\boldsymbol{\alpha}$ matrix elements, allowing for more general cases.

Relating cosmological parameters to information loss coefficients is a crucial aspect of this analysis. The starting point is the Fisher Information Matrix (FIM). For cosmological parameters $\theta$, the FIM elements are given by:
\begin{equation}
F_{ij} = \left\langle \frac{\partial^2 L}{\partial\theta_i \partial\theta_j} \right\rangle
\end{equation}
where $L$ is the log-likelihood of the data given the parameters. The inverse of the FIM provides a lower bound on the covariance matrix of the parameter estimates (Cramer--Rao bound). This gives us an idea of the best possible constraints on cosmological parameters. For a combined analysis, we construct a total Fisher matrix:
\begin{equation}
F_{total} = F_{CMB} + F_{BAO} + F_{SN} + F_{H(z)} + ... = \sum F_i,
\end{equation}
where each $F_i$ is the Fisher matrix for the respective probe. We can now introduce loss coefficients for each probe and parameter:
\begin{equation}
F'_{total} = \alpha_{CMB} \circ F_{CMB} + \alpha_{BAO} \circ F_{BAO} + \alpha_{SN} \circ F_{SN} + \alpha_{H(z)} \circ F_{H(z)} + ...
\end{equation}
{where $\circ$ denotes the Hadamard product (element-wise). This element-wise multiplication is specifically chosen because the matrices $\boldsymbol{\alpha}_i$ are designed to act as ``attenuation masks'' on the original Fisher matrices. Each element $\alpha_{ij}$ within $\boldsymbol{\alpha}_i$ directly scales the corresponding $ij$-th element of the Fisher matrix $F_i$, allowing for a granular, heterogeneous reduction of information across different parameter correlations. This contrasts with standard matrix multiplication, which would imply a linear transformation of the parameter space, and is not what our model intends to represent for information loss.}

This approach allows us to maximize the information content from diverse observational probes while quantifying and accounting for potential information loss. By carefully modeling the interplay between different datasets and their associated systematics, we can obtain robust constraints on cosmological parameters and gain insights into the nature of information degradation in cosmology. In particular, the Hubble tension refers to the discrepancy between measurements of the Hubble constant ($H_0$) derived from early Universe probes and those from late Universe probes. Understanding this division is crucial for analyzing the tension and applying our Fisher matrix approach with information \mbox{loss coefficients}.

\subsection{Generation of Information Loss Coefficients}
The likelihood function is modified to include the information loss coefficients and the priors are defined for $\alpha_{ij}$ from the beta distributions, Beta(a,b), where a and b are hyperparameters controlling the shape of the distribution. The coefficients $\alpha_{ij}$ represent the fraction of theoretical information effectively extracted from the observations. They range from 0 to 1, where 1 means no information loss, and 0 means total information loss. We use a Markov chain Monte Carlo (MCMC) algorithm (e.g., Metropolis--Hastings or Hamiltonian Monte Carlo) to estimate the $\alpha_{ij}$ values. The MCMC explores the parameter space, including both cosmological parameters and $\alpha_{ij}$.

The sampling process follows two basic steps. First, we start with initial values for $\alpha_{ij}$ (can be prior values or arbitrary values between 0 and 1). Then, at each MCMC iteration:
\begin{itemize}
    \item Propose new values for $\alpha_{ij}$.
    \item Calculate the modified Fisher matrix $F'_{ij} = \alpha_{ij} \cdot F_{ij}$.
    \item Calculate the likelihood using this modified matrix.
    \item Accept or reject new values according to the MCMC acceptance ratio.
\end{itemize}

It is important to note that the likelihood function is modified to include $\alpha_{ij}$:
\begin{equation}
L(\theta,\alpha|\text{data}) \propto \exp\left[-0.5(\theta-\theta_{fid})^T F' (\theta-\theta_{fid})\right] \cdot p(\alpha)
\end{equation}
where $\theta$ are the cosmological parameters, $\theta_{fid}$ are fiducial values, $F'$ is the modified Fisher matrix, and $p(\alpha)$ is the prior to $\alpha_{ij}$. After MCMC, we analyze the posterior distribution of $\alpha_{ij}$ and calculate means, medians, and confidence intervals for each $\alpha_{ij}$.

\subsection{Separating Measurement Error and Information Loss in Hubble Tension}
\label{sec:3.2}
{We propose a Bayesian hierarchical model to separate the contributions of measurement error and information loss to the observed Hubble tension, as defined in Equation (1).}
 For the purpose of illustrating our methodology, in this work we focus on the well-known tension between the CMB and SNIa measurements of $H_0$. Future work will extend this framework to incorporate a broader range of cosmological datasets.

The variance inflation factors $\lambda_{CMB}$ and $\lambda_{SNIa}$ are defined as:
\begin{equation}
\lambda_{CMB} = \frac{Var(H_{0,CMB})_{total}}{Var(H_{0,CMB})_{marginal}}
\end{equation}
\begin{equation}
\lambda_{SNIa} = \frac{Var(H_{0,SNIa})_{total}}{Var(H_{0,SNIa})_{marginal}}
\end{equation} \\
\noindent These equations rigorously define the variance inflation factor for CMB and SNIa measurements. The functional form, as a simple ratio of variances, directly quantifies the extent to which the total uncertainty ($\text{Var}(H_{0,\text{SN Ia}})_{\text{total}}$) exceeds the uncertainty attributed solely to conventional measurement errors ($\text{Var}(H_{0,\text{SN Ia}})_{\text{marginal}}$). This empirical definition captures any additional variance, including that arising from information loss due to parameter marginalization or unmodeled systematic effects, that is not captured by the reported measurement uncertainty. In other words,
 the variance inflation factors quantify how much the variance of $H_0$ increases when we account for the full parameter space compared to considering $H_0$ in isolation.

The information loss component ($E_i$) is calculated as the standard deviation of the additional variance not accounted for by measurement errors, consistent with the rigorous separation in Section \ref{sec:2.4}:
\begin{equation}
E_i = \sqrt{(\lambda_{CMB} - 1)\sigma_{CMB,obs}^2 + (\lambda_{SNIa} - 1)\sigma_{SNIa,obs}^2}
\end{equation}

\noindent This equation calculates the standard deviation of the information loss component $E_i$. Its functional form is derived directly from the variance decomposition presented in Section \ref{sec:2.4}. The term $(\lambda - 1)\sigma_{\text{obs}}^2$ represents the additional variance beyond the reported measurement error that is attributable to information loss for each probe. By summing these additional variance contributions for CMB and SNIa and taking the square root, we obtain the combined standard deviation of the information loss. This quadrature sum is appropriate, as it assumes that the information loss contributions from the two distinct probes are independent sources of variance, thereby adding linearly in terms of squared standard deviations.

The real tension $T_{real}$ is then estimated as a parameter within the Bayesian framework, representing the true underlying discrepancy.
Consistent with the observational data outlined in Section \ref{sec:1}, our analysis uses the following key measurements of the Hubble constant:

\begin{itemize}[label=,leftmargin=0em,labelsep=4mm]
\item[] $H_{0,CMB} = 67.24 \pm 0.35$ km/s/Mpc from Planck 2018 data.
\item[] $H_{0,SNIa} = 73.17 \pm 0.86$ km/s/Mpc from the SH0ES (Supernova $H_0$ for the Equation of State) team.
\end{itemize}

The combination of these values yields an observed tension of $T_{observed} = 5.94$ km/s/Mpc. Our aim is to verify how much uncertainty and loss of information contribute to this value.

\subsection{Model Specification}
Bayesian analysis provides a natural framework for decomposing the Hubble tension, allowing for the explicit incorporation of uncertainties at multiple levels of the problem and direct quantification of the relative contributions from different sources of variance. Our Bayesian hierarchical model formally characterizes the observed tension $H_0$ between the CMB and SNIa datasets as arising from three fundamental components: the real physical tension ($T_{real}$), measurement error ($E_m$), and information loss due to projection from multidimensional parameter spaces ($E_i$). The hierarchical structure of the model recognizes that these components are not directly observed but emerge from a generative process involving variance inflation factors for each method ($\lambda_{CMB}$ and $\lambda_{SNIa}$).

At the top level of the hierarchy, we model the observed difference between the estimates $H_0$ as a quantity arising from an underlying generative process. At the intermediate level, this difference is decomposed into components with distinct physical and statistical meanings. At the lower level, the variance inflation factors, which capture the covariance structure of the full parameter spaces, are treated as parameters to be estimated from the data, rather than fixed values. A crucial advantage of this approach is that uncertainty in the estimation of variance inflation factors is automatically propagated to our conclusions about the magnitude of the real tension. Additionally, the Bayesian model allows for a direct interpretation of confidence intervals for the real tension and provides a foundation for formal model comparisons should different structures for tension decomposition \mbox{be considered.}

The complete mathematical formulation of the model begins with the specification of prior distributions for all parameters. These priors represent our knowledge or beliefs about the parameters before observing the specific data in this analysis. The choice of these prior distributions is guided by both theoretical considerations and previous empirical results related to the structure of cosmological parameter spaces.

\subsubsection{Priors}

{
The choice of prior distributions is a critical step in Bayesian analysis, as they encode our prior knowledge or assumptions about the parameters before observing the data. We carefully selected our priors to balance weak informativeness with physical consistency:}

{ For the true values of the Hubble constant, $H_{0,\text{CMB},\text{true}}$  and $H_{0,\text{SN Ia},\text{true}}$, we employ Normal distributions ($N$):}
\begin{align}
H_{0,\text{CMB},\text{true}} &\sim N(\mu_{\text{CMB}}, \sigma_{\text{CMB}}) \label{eq:27}\\
H_{0,\text{SN Ia},\text{true}} &\sim N(\mu_{\text{SN Ia}}, \sigma_{\text{SN Ia}}) \label{eq:28}
\end{align}
\noindent This choice is standard for continuous parameters that are expected to be centered on a specific value and possess a quantifiable uncertainty. The Normal distribution is maximal entropy given a mean and variance, making it a flexible choice. We set these as weakly informative priors by centering them on the observed values of $H_0$ (\mbox{e.g., $\mu_{\text{CMB}} = 67.24$ km/s/Mpc}) but assigning sufficiently wide standard deviations (e.g., $\sigma_{\text{CMB}}$ much larger than the observed uncertainty) to ensure that the data primarily drives the posterior inference, rather than the prior. This approach allows the MCMC sampler to efficiently explore the parameter space while maintaining physical realism.

\noindent For the information loss coefficients, $\alpha_{\text{CMB}}$  and $\alpha_{\text{SN Ia}}$, we use Beta distributions:
\begin{align}
\alpha_{\text{CMB}} &\sim \text{Beta}(a_{\text{CMB}}, b_{\text{CMB}}) \label{eq:29}\\
\alpha_{\text{SN Ia}} &\sim \text{Beta}(a_{\text{SN Ia}}, b_{\text{SN Ia}}) \label{eq:30}
\end{align}
\noindent The Beta distribution is a continuous probability distribution defined on the interval $[0, 1]$, which makes it the natural and most appropriate choice for parameters that represent proportions, fractions, or probabilities. Since the coefficients $\alpha$ represent the ``fraction of theoretical information effectively extracted from observations'' (i.e., $\alpha \in [0, 1]$), the Beta distribution is perfectly suited. Its two positive shape parameters, $a$ and $b$, provide significant flexibility to model various prior beliefs about the distribution of information loss, ranging from uniform (e.g., Beta (1,1)) to skewed toward higher (e.g., Beta (5,1)) or lower (e.g., Beta (1,5)) values. This flexibility allows us to specify informative but relatively wide priors, as discussed in detail in the Bayesian implementation section (Section \ref{sec:3.5}), allowing the data to primarily inform the magnitude of information loss.


\subsubsection{Likelihood}
The likelihood function is given by:
\begin{align}
\label{eq:likelihood_definition} 
L(&H_{true,CMB},H_{true,SNIa},\alpha_{CMB},\alpha_{SNIa}|H_{obs,CMB},H_{obs,SNIa}) \nonumber \\
&= p(H_{obs,CMB},H_{obs,SNIa}| H_{true,CMB},H_{true,SNIa},\alpha_{CMB},\alpha_{SNIa})
\end{align}
with the expressions below representing the generative model for our observations, that is, how the observed measurements of $H_0$ are generated as a function of the true values and their respective uncertainties.
\begin{equation}
H_{0,CMB,obs} \sim \mathcal{N}(H_{0,CMB,true},\sigma_{CMB,obs}^2 / \alpha_{CMB})
\end{equation}
\begin{equation}
H_{0,SNIa,obs} \sim \mathcal{N}(H_{0,SNIa,true},\sigma_{SNIa,obs}^2 / \alpha_{SNIa})
\end{equation} 

These equations specify the generative model for the observed CMB and SNIa measurements of $H_0$. The use of a Normal (Gaussian) distribution is a standard assumption for modeling observational errors, often justified by the Central Limit Theorem when multiple small error sources contribute to the total uncertainty. 
The important aspect of its functional form lies in the variance term which explicitly incorporates the information loss coefficients $\alpha_{\text{CMB}}$ and
 $\alpha_{\text{SNIa}}$ into the precision of the observed data. Since $\alpha$ is a fraction ($0 < \alpha \le 1$), dividing by it effectively inflates the reported variances $\sigma_{\text{CMB,obs}}^2$ and $\sigma_{\text{SNIa,obs}}^2$. A smaller $\alpha$ (indicating greater information loss) leads to a larger variance, implying a less precise observed measurement. This directly reflects how the ``loss of information'' manifests in the observational data, making the observed $H_0$ value effectively less constrained.

Expanding these expressions using probabilistic models, we have the following:
\begin{align}
L &= \frac{1}{\sqrt{2\pi(\sigma_{CMB,obs}^2 / \alpha_{CMB})}} \exp\left[ -\frac{(H_{obs,CMB} - H_{true,CMB})^2}{2(\sigma_{CMB,obs}^2 / \alpha_{CMB})} \right] \times \nonumber \\
  &\quad \frac{1}{\sqrt{2\pi(\sigma_{obs,SNIa}^2 / \alpha_{SNIa})}} \exp\left[ -\frac{(H_{obs,SNIa} - H_{true,SNIa})^2}{2(\sigma_{obs,SNIa}^2 / \alpha_{SNIa})} \right]
\end{align}

\subsection{Derived Quantities} 
\label{sec:3.4}
\begin{equation}
T = H_{0,SNIa,obs} - H_{0,CMB,obs}
\end{equation}
\begin{equation}
E_m = \sqrt{\sigma_{CMB,obs}^2 + \sigma_{SNIa,obs}^2}
\end{equation}
\begin{equation}
E_i = \sqrt{(\lambda_{SNIa} - 1)\sigma_{SNIa,obs}^2 + (\lambda_{CMB} - 1)\sigma_{CMB,obs}^2}
\end{equation}
\begin{equation}
T_{real} = H_{0,SNIa,true} - H_{0,CMB,true}
\end{equation}


\subsection{Bayesian Implementation}
\label{sec:3.5}
The Bayesian analysis was implemented using the Stan probabilistic programming language, which employs a No-U-Turn Sampler (NUTS), a highly efficient variant of Hamiltonian Monte Carlo (HMC). This choice is particularly advantageous for complex, high-dimensional models as it navigates the parameter space effectively, reducing issues like random walk behavior and improving sampling efficiency.

We ran 4 independent MCMC chains, each initialized from different random starting points to ensure robust exploration of the posterior distribution and to facilitate convergence diagnostics. Each chain consisted of 5000 iterations. The first 2000 iterations of each chain were designated as warm-up (or burn-in) and discarded. This warm-up phase allows the sampler to adapt its parameters and converge to the target posterior distribution, ensuring that the subsequent samples are representative. This configuration resulted in a total of 12,000 effective posterior samples (4 chains $\times$ (5000 $-$ 2000) samples/chain) used for subsequent inference.

The prior distributions for all the model parameters were carefully specified. For the true Hubble constant values ($H_{0,CMB,true}$ and $H_{0,SNIa,true}$), we used weakly informative normal priors, centered on the observed values, but with sufficiently wide standard deviations to allow the data to primarily drive the inference. For the information loss coefficients ($\alpha_{CMB}$ and $\alpha_{SNIa}$), Beta priors were used. These were set to be informative but relatively wide, reflecting the expectation that some information loss might occur, but avoiding overly strong assumptions about its magnitude. This choice aligns with the goal of allowing the data to inform the extent of information loss, while respecting the physical bounds of the parameters $\alpha$ (0 to 1). The prior for the real tension component ($T_{real}$) was chosen to be non-informative (e.g., a very wide normal distribution or a flat prior), ensuring that the data primarily drive its posterior distribution.

The convergence of the MCMC chains was rigorously assessed to ensure that the samples accurately represent the target posterior distribution. We monitored the R-hat statistic for all parameters, aiming for values close to 1.0 (typically below 1.01--1.05), which indicates good mixing and convergence across chains. Furthermore, the effective sample size (ESS) was checked to ensure a sufficient number of independent samples for reliable inference, generally targeting ESS values greater than 400--1000 for each parameter. These diagnostics confirmed that the chains had adequately explored the parameter space.

From the converged posterior samples, summary statistics (mean, median, standard deviation) were calculated for all model parameters and derived quantities (such as $E_m$, $E_i$, and $T_{real}$). Posterior confidence intervals (e.g., 95\% confidence intervals) were calculated directly from the percentiles of the MCMC samples, providing a robust measure of uncertainty for each parameter. The full Stan model code and data used for this analysis are available upon request/in a supplementary repository to ensure reproducibility.

\section{Results and Interpretation}
Our Bayesian analysis, employing the framework detailed in Sections \ref{sec:2}  and \ref{sec:3}, provides a robust decomposition of the observed Hubble tension. The primary objective is to quantify the contributions of the real physical discrepancy, measurement errors, and loss of information to the overall tension. The results strongly indicate that the Hubble tension remains a statistically significant phenomenon even after accounting for these factors.

\subsection{Posterior Estimates}

\begin{figure}[H]
\includegraphics[width=115mm]{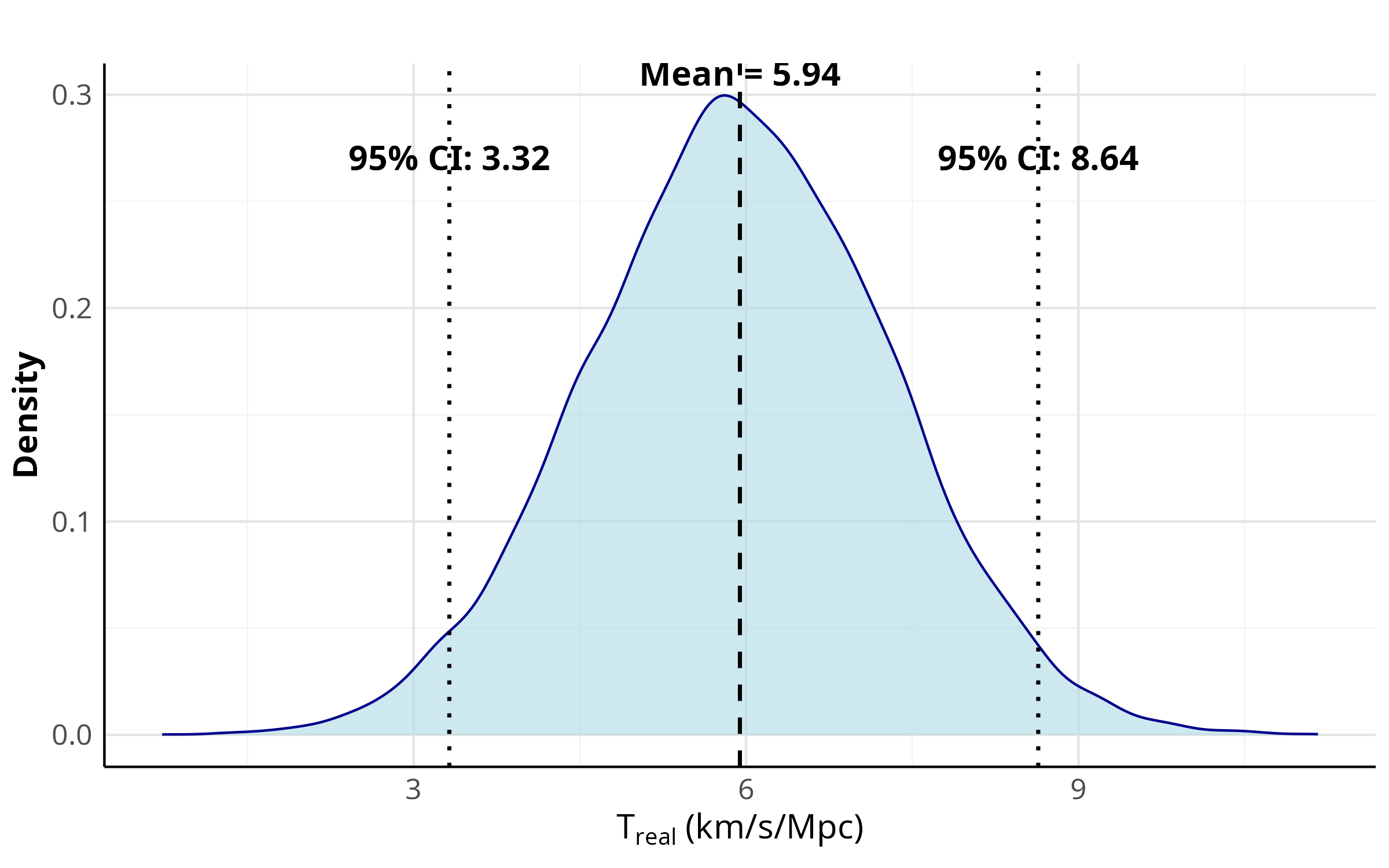}
\vspace{-0.32cm}
\caption{Posterior distribution of $T_{real}$. The vertical dashed line indicates the mean value, while the dotted vertical lines indicate the 95\% confidence interval.}
\label{bayes_posterior}
\end{figure}

\begin{figure}[H]
\includegraphics[width=115mm]{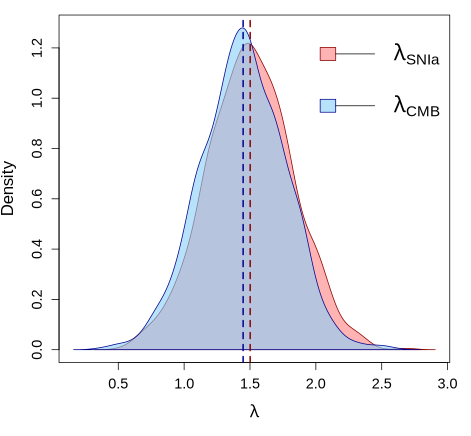}

\caption{Posterior distribution of $\lambda_{CMB}$ and $\lambda_{SNIa}$. Dashed lines indicate the respective mean values.}
\label{fig2}

\end{figure}

\begin{itemize}
    \item {Real Tension ($T_{real}$):} The posterior distribution of $T_{real}$ shows a mean value of \mbox{5.94 km/s/Mpc} and a median of 5.92 km/s/Mpc. The 95\% confidence interval is [3.32, 8.64] km/s/Mpc. As illustrated in Figure \ref{bayes_posterior}, this interval clearly excludes zero, indicating that the observed discrepancy is not merely a statistical artifact, but reflects a genuine physical phenomenon. The Bayesian significance further supports this, with 100\% of posterior samples for $T_{real}$ greater than zero.

    \item {Variance Inflation Factors ($\lambda$):} The estimated variance inflation factors are $\lambda_{CMB} = 1.45$ (95\% CI: [0.81, 2.05]) and $\lambda_{SNIa} = 1.50$ (95\% CI: [0.85, 2.12]). These values, also visualized in Figure \ref{fig2}, are significantly greater than 1, confirming the presence of additional variance beyond the reported measurement uncertainties. The similarity in the values of $\lambda_{CMB}$ and $\lambda_{SNIa}$ suggests that both the CMB and SNIa measurements are affected by comparable levels of information loss or unmodeled systematic uncertainties.
\end{itemize}

\vspace{-12pt}
\vspace{-9pt}

\subsection{Decomposition of Observed Tension}

A critical aspect of our analysis is the decomposition of the observed Hubble tension into its constituent components. This decomposition is based on the posterior estimates of our model's fundamental parameters and their derived quantities.
{We clarify that the components---Real Tension ($T_{\text{real}}$), Measurement Error ($E_m$), and Information Loss ($E_i$)---are derived quantities from the MCMC posterior samples, not directly sampled independent parameters. For each MCMC sample drawn from the joint posterior distribution of our primary parameters ($H_{0,\text{CMB},\text{true}}$, $H_{0,\text{SN Ia},\text{true}}$, $\alpha_{\text{CMB}}$, and $\alpha_{\text{SN Ia}}$), we compute the corresponding values for $T_{\text{real}}$, $E_m$, and $E_i$ using the definitions in Section \ref{sec:3.4}. This process generates full posterior distributions for each of these derived quantities.

To quantify their relative contributions to the total observed tension variance, we calculate the variance of the posterior distribution of each derived component: $\text{Var}(T_{\text{real}})$, $\text{Var}(E_m)$, and $\text{Var}(E_i)$. The total variance used for this decomposition is then taken as the sum of these component variances, implicitly treating their contributions as orthogonal for the purpose of this analysis: $\text{Var}(T_{\text{observed}}) \approx \text{Var}(T_{\text{real}}) + \text{Var}(E_m) + \text{Var}(E_i)$. This decomposition is presented in Figure \ref{fig3}.

\begin{itemize}
    \item {Real Tension:} Approximately 77.78\% of the variance of the observed tension is attributed to the real physical discrepancy ($T_{real}$). This is the dominant component, reinforcing the conclusion that the Hubble tension is primarily a genuine astrophysical puzzle.

    \item {Measurement Error:} In total, 12.98\% of the variance of the observed tension is accounted for by standard measurement errors ($E_m$).

    \item {Information Loss:} The remaining 9.24\% of the variance of the observed tension is attributed to information loss ($E_i$), arising from model simplifications and projection of the parameter space. This component, while smaller than the real tension, is non-negligible and highlights the importance of accounting for such effects in cosmological analyses.
\end{itemize}
\vspace{-17pt}
\begin{figure}[H]

\includegraphics[width=95mm]{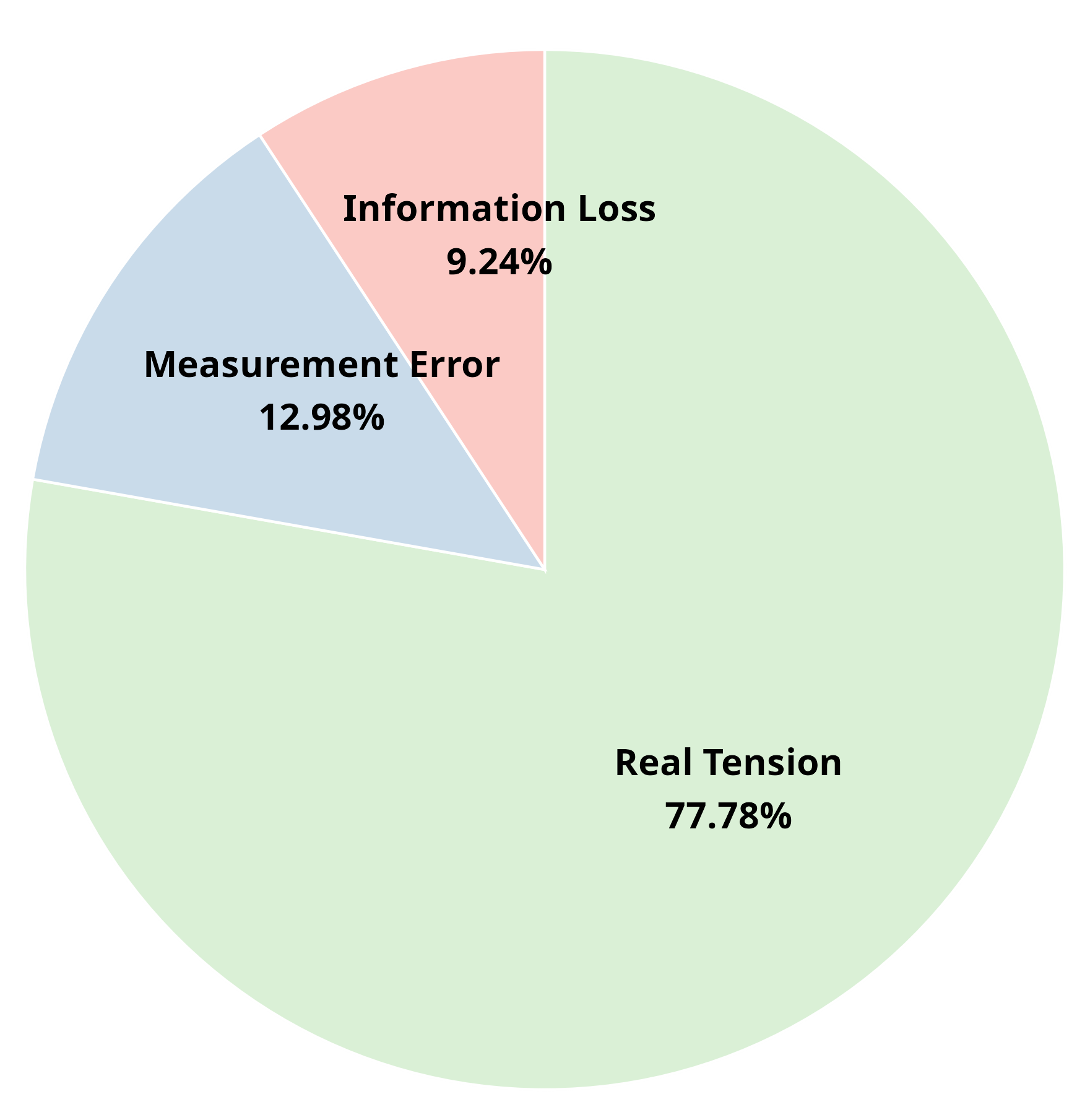}
\vspace{-0.2cm}
\caption{Percentage contributions of real tension, measurement error, and information loss to the total variance of the observed tension.}
\label{fig3}

\end{figure}

These findings suggest that the Hubble tension persists even when accounting for measurement errors and information loss. The discrepancy between the CMB and SNIa measurements of $H_0$ appears to be largely real and cannot be explained solely by measurement uncertainties or information loss. This implies that the Hubble tension may indeed point to new physics beyond the standard cosmological model or to unidentified systematic errors in one or both measurement methods.

This result is complementary to those shown previously.
In Figure \ref{bayes_posterior} we present the
posterior distribution of real tension ($T_{real}$), clearly showing that $T_{real}$ is significantly non-zero.  In Figure \ref{fig2} we show
the posterior distributions of the variation inflation factors ($\lambda_{CMB}$ and $\lambda_{SNIa}$). This allows for a visual comparison of the estimated information loss for each probe and confirms that these factors are greater than 1. In Figure
\ref{fig3}, we present a pie chart that illustrates the
decomposition of observed tension variance, with the percentage contributions of real tension, measurement error, and information loss to the total variance of observed tension. 

To gain a deeper understanding of the interplay between these components and the underlying parameters driving this decomposition, we present the corner plot of the simulated posterior distributions in Figure \ref{fig4}. This visualization allows for a qualitative assessment of the marginal distributions of each parameter ($T_{real}$, $\lambda_{CMB}$, $\lambda_{SNIa}$, $E_i$, $H_{0,CMB,true}$, $H_{0,SNIa,true}$) as well as their joint dependencies.

The diagonal panels of Figure \ref{fig4} show the marginal probability density functions. For $T_{real}$, the distribution is clearly centered at a non-zero value, reinforcing the conclusion from Figure \ref{fig3} that a significant portion of the observed tension stems from a genuine physical phenomenon. In particular, the posterior distributions for the variance inflation factors, $\lambda_{CMB}$ and $\lambda_{SNIa}$, exhibit a distinct left-truncation at 1. This characteristic is consistent with our model's theoretical premise, where $\lambda \geq 1$ implies variance inflation due to information loss and not an overestimation of uncertainties, as discussed in Section \ref{sec:3.2}}.

The off-diagonal panels in the upper triangle of Figure \ref{fig4} illustrate the bivariate correlations between the parameters. The use of hexagonal binning effectively visualizes the density of the simulated samples, providing clear insight into regions of high probability density. We observe strong positive correlations between parameters whose relationships are directly defined within the model. Specifically, there is a clear positive correlation between $T_{real}$ and $H_{0,SNIa,true}$, consistent with $T_{real}$ representing the difference between $H_{0,SNIa,true}$ and $H_{0,CMB,true}$. Similarly, a strong positive correlation is evident between the variance inflation factors ($\lambda_{CMB}$ and $\lambda_{SNIa}$) and the derived information loss component ($E_i$), which is a direct consequence of the mathematical formulation of $E_i$' as a function of these inflation factors. This strong interdependency is directly aligned with the mathematical formulation where $E_i$ is a function of these inflation factors, providing distributional evidence for the 9.24\% contribution to information loss highlighted in Figure \ref{fig3}. 

In addition, we quantify the contributions of each component to the observed Hubble tension in terms of standard deviations ($\sigma$). The observed tension between the $H_0$ values of Planck CMB ($67.24 \pm 0.35$ km/s/Mpc) and SH0ES SNIa ($73.17 \pm 0.86$ km/s/Mpc) is $T_{observed} = 5.94$ km/s/Mpc. The combined uncertainty of this observed tension is $\sqrt{0.24^2 + 0.86^2} \approx 0.89$ km/s/Mpc. Therefore, the observed Hubble tension corresponds to approximately $6.39\sigma$.

\begin{figure}[H]

\includegraphics[width=120mm]{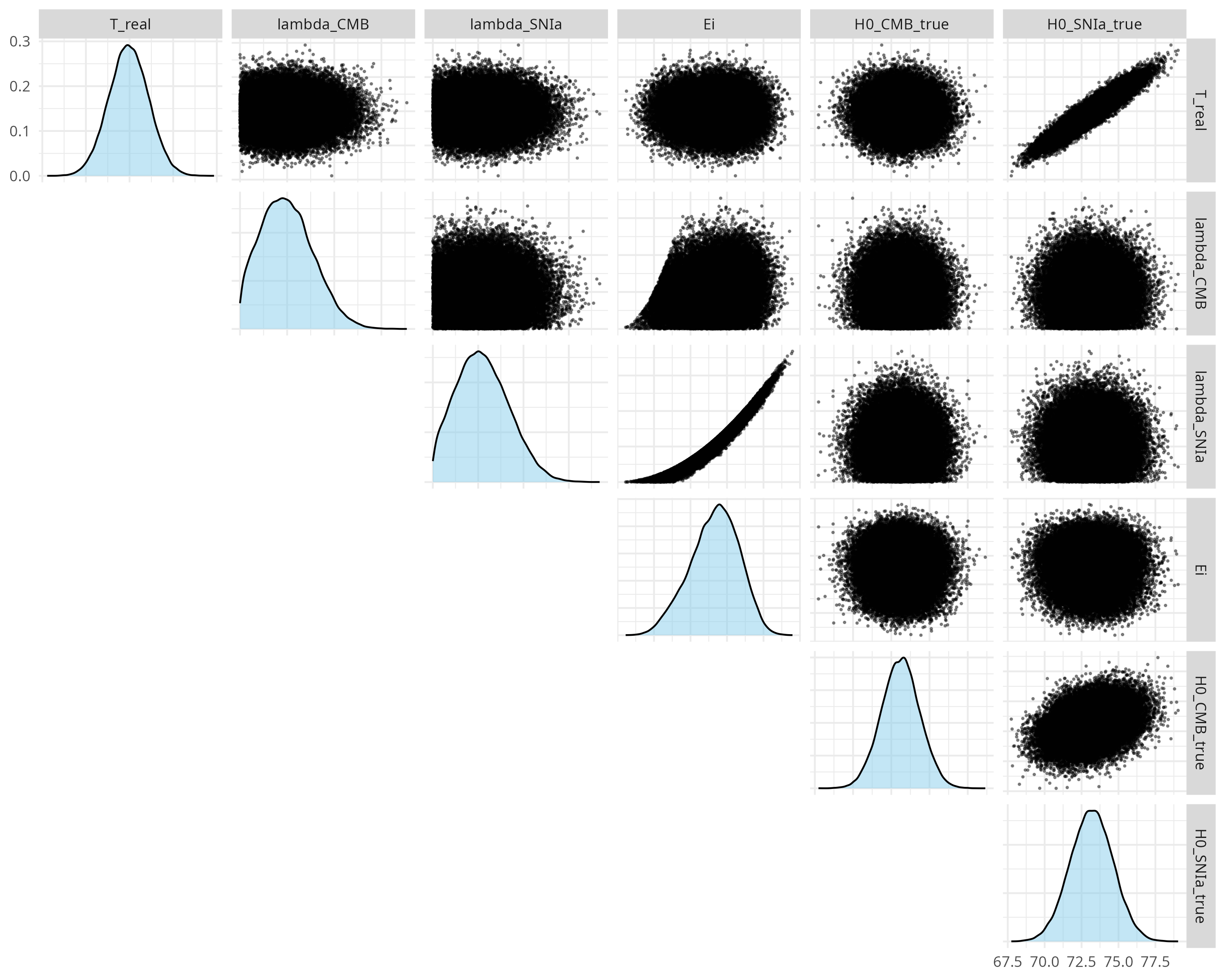}
\vspace{-0.25cm}
\caption{Corner plot depicting the approximate joint and marginal posterior distributions of the key parameters in our simulated Bayesian hierarchical model for the Hubble tension decomposition.}
\label{fig4}

\end{figure}

Building upon the variance decomposition presented previously, we can translate these contributions into their respective magnitudes in terms of sigma:
\begin{itemize}
    \item {Real Tension:} Accounting for 77.78\% of the variance, corresponds to $\sqrt{0.7778} \times 6.39\sigma \approx 5.64\sigma$ of the observed tension.
    \item {Measurement Error:} Contributing 13.98\% to the variance, standard measurement errors account for approximately $\sqrt{0.1398} \times 6.39\sigma \approx 2.39\sigma$ of the observed tension.
    \item {Information Loss:} With 9.24\% of the variance attributed to information loss, this component contributes approximately $\sqrt{0.0924} \times 6.39\sigma \approx 1.94\sigma$ to the observed tension.
\end{itemize}

This sigma-based quantification reinforces our conclusion that the Hubble tension is predominantly a real physical phenomenon, with a substantial portion of its magnitude stemming from a genuine discrepancy that cannot be fully explained by statistical or information loss effects alone.

\section{Discussion}

Our analysis provides compelling evidence that the Hubble tension remains significant even after accounting for both measurement errors and information loss due to parameter-space projection. The posterior distribution of the real tension component ($T_{real}$) shows a mean value of 5.94 km/s/Mpc with a confidence interval of 95\% of [3.32, 8.64] km/s/Mpc, clearly excluding zero from the range of plausible values. {It is important to note that $T_{\text{real}}$ represents the discrepancy that persists after rigorously accounting for the uncertainties of the statistical measurement and the inherent loss of information from the marginalization of the parameters within the standard cosmological model framework. Therefore, a significant $T_{\text{real}}$ strongly suggests that the observed discrepancy between CMB and SNIa measurements likely reflects a genuine physical phenomenon beyond what can be explained by these statistical and information-processing effects alone. This finding does not preclude the possibility that new physics beyond $\Lambda$CDM could ultimately resolve the tension; rather, it provides robust statistical evidence that such a resolution would indeed require going beyond the standard model's current observational and data analysis interpretations.}
This finding aligns with the prevailing consensus in the cosmological community that the Hubble tension is a robust discrepancy, often reported at a significance level exceeding 4--5$\sigma$ in various independent analyzes \cite{10.3847/2041-8213/ac5c5b, 10.1093/mnras/stad3724, 10.1016/j.physletb.2024.138717, 10.1140/epjc/s10052-024-12999-w}. Our work reinforces this by demonstrating that even when explicitly modeling and quantifying potential sources of uncertainty and information degradation, the core tension persists.

A key distinguishing feature of our approach, based on previous Bayesian treatments of cosmological tensions \cite{10.1093/mnras/sty418, 10.1093/mnras/sty3082, 10.1103/PhysRevD.103.L041301}, is the explicit parameterization and decomposition of the observed tension into three distinct components: standard measurement errors, information loss errors arising from model simplifications and parameter-space projection, and the real physical tension. For example, while Feeney et al. \cite{10.1093/mnras/sty418} applied a Bayesian hierarchical model to clarify tension within the local distance ladder, our framework extends this by specifically isolating and quantifying the contribution of information loss ($E_i$) through the introduction of variance inflation factors ($\lambda$). Our results indicate that approximately 78\% of the variance of the observed tension is attributed to the real tension, the measurement error that accounts for 14\% and the loss of information for the remaining 9\%. The estimated variance inflation factors ($\lambda_{CMB} \approx 1.45$ and $\lambda_{SNIa} \approx 1.50$) being significantly greater than unity further underscore the importance of accounting for these effects, as they suggest that reported uncertainties might be underestimated or that parameter correlations lead to non-trivial information loss when marginalizing. This figure, therefore, represents the magnitude of the tension remaining after explicitly modeling identifiable error sources; while strongly suggestive of new physics, it is formally an upper bound for direct evidence of such physics, as it inherently absorbs any unquantified systematic effects not accounted for by our $\lambda$ parameters.
This decomposition provides a more nuanced understanding of the tension's origins, allowing us to confidently assert that the majority of the discrepancy is not an artifact of our analytical pipeline.

Future studies should extend our variance decomposition framework to incorporate a broader observational landscape. Particularly valuable would be the inclusion of probes that sample intermediate redshifts between the recombination epoch probed by the CMB and the relatively local universe explored by SNIa. The addition of such complementary datasets is expected to have a significant impact on the information loss component. By breaking existing parameter degeneracies, these new data sources should reduce the correlations between $H_0$ and other cosmological parameters. Consequently, our methodology would predict a decrease in the information loss component ($E_i$) and a convergence of the variance inflation factors ($\lambda$) closer to unity. This reduction in information loss would further strengthen the robustness of the real tension, should it persist, by demonstrating that it is not an artifact of unresolved correlations or projection effects. Conversely, if the information loss component were to remain substantial or even increase with the inclusion of more data, it would point toward more complex or yet unidentified systematic issues in the combined dataset analysis.  Furthermore, a multiprobe analysis would allow cross-validation of the inflation factors of the estimated variance. If similar values of $\lambda$ emerge from independent dataset combinations, this would strengthen confidence in our quantification of information loss. In contrast, significant variations in these factors across different probe combinations might indicate probe-specific systematics or modeling assumptions that warrant further investigation.

Although our current results indicate that the Hubble tension remains robust even after accounting for information loss, we cannot exclude the possibility that more complex projection effects, perhaps involving higher-order moments of parameter distributions or nonlinear parameter degeneracies, might emerge in a more diverse dataset combination. The Bayesian hierarchical framework we have developed is well-suited for such extensions, as it naturally accommodates additional complexity through appropriate prior specifications and model comparison tools.

In conclusion, while our analysis provides important insights into the nature of the Hubble tension using two foundational cosmological probes, a definitive assessment of the contribution of information loss to this tension will require a more comprehensive observational foundation. This represents a promising direction for future research that could further illuminate whether the Hubble tension ultimately demands new physics beyond the standard cosmological model, as discussed in recent works
\cite{10.1016/j.newar.2022.101659,10.1016/j.jheap.2022.04.002,10.3390/universe9020094}.

\vspace{6pt} 





\authorcontributions{ A. L. B. Ribeiro was specifically responsible for the Conceptualization, Resources, and Supervision of the work. N. M. N. da Rocha undertook the tasks of Methodology, Project administration, and Visualization. F. B. S. Oliveira's contributions include Software development and Validation. Furthermore, all listed authors collectively engaged in critical aspects such as Data curation, Formal analysis, Writing - original draft, and Writing - review and editing for the manuscript. }

\funding{This research did not receive any external funding.}

\dataavailability{This research did not use direct observational data; instead, it employed cosmological parameter values as reported in the published literature.}

\acknowledgments{We thank the anonymous reviewers for their helpful contributions to this paper. NMNR thanks the support of PROBOL-UESC.
ALBR thanks the support of CNPq, grant 316317/2021-7.}

\conflictsofinterest{The authors declare no conflicts of interest.} 



\abbreviations{Abbreviations}{
The following abbreviations are used in this manuscript:
\\

\noindent 
\begin{tabular}{@{}ll}
$H_0$ & The Hubble Constant\\
$\Lambda$CDM & Lambda Cold Dark Matter\\
CMB & Cosmic Microwave Background\\
SNIa & Type Ia supernovae \\
BAO & Baryon Acoustic Oscillations \\
TRGB & Tip of the Red Giant Branch \\
JWST & The James Webb Space Telescope \\
KL & Kullback--Leibler \\
FIM & Fisher Information Matrix \\
IL & Information Loss \\
TCI & Total Constraining information \\
MCMC & Markov Chain Monte Carlo \\
$\mathcal{N}$ & Normal Distributions \\
NUTS & No-U-Turn \\
HMC & Hamiltonian Monte Carlo \\
\end{tabular}
}


\begin{adjustwidth}{-\extralength}{0cm}

\reftitle{References}



\begin{thebibliography}{999}

\bibitem[Freedman, W. L. and Madore, B. F.(2023)]{10.1088/1475-7516/2023/11/050}
Freedman, W.L.; Madore, B.F. Progress in direct measurements of the Hubble constant. {\em J. Cosmol. Astropart. Phys.} {\bf 2023}, {\em 11}, 050.
\bibitem[Verde, L. and Schöneberg, N. and Gil-Marín, H.(2024)]{10.1146/annurev-astro-052622-033813}
Verde, L.; Schöneberg, N.; Gil-Marín, H. A Tale of Many $H_0$. {\em Annu. Rev. Astron. Astrophys.} {\bf 2024}, {\em 62}, 287-331.

\bibitem[Hu, J. P., and Wang, F. Y.]
{10.3390/universe9020094} Hu, J.P.;  Wang, F.Y.
Hubble tension: The evidence of new physics.
{\em Universe} {\bf 2023}, {\em 9}, 94.

\bibitem[ Perivolaropoulos, L., and Skara, F.]
{10.1016/j.newar.2022.101659} Perivolaropoulos, L.; Skara, F. Challenges for $\Lambda$CDM: An update
{\em New Astron. Rev.} {\bf 2022},
{\em 95}, 101659.

\bibitem[Riess, A. G. et al.(2022)]{10.3847/2041-8213/ac5c5b}
Riess, A.G.; Yuan, W.; Macri, L.M.; Scolnic, D.; Brout, D.; Casertano, S.;  Zheng, W.  A Comprehensive Measurement of the Local Value of the Hubble Constant with 1 km s\textsuperscript{{−1}} Mpc\textsuperscript{{−1}} Uncertainty from the Hubble Space Telescope and the SH0ES Team. {\em  Astrophys. J. Lett.} {\bf 2022}, {\em 934}, L7.

\bibitem[Wang, B. and L{\'o}pez-Corredoira, M. and Wei, J.(2024)]{10.1093/mnras/stad3724}
Wang, B.; L{\'o}pez-Corredoira, M.; Wei, J. The Hubble tension survey: A statistical analysis of the 2012--2022 measurements. {\em Mon. Not. R. Astron. Soc.} {\bf 2024}, {\em 527}, 7692--7700.

\bibitem[Abdalla, E., Abellán, G. F., Aboubrahim, A., Agnello, A., Akarsu, Ö., Akrami, Y., ... \& Pettorino, V.]{10.1016/j.jheap.2022.04.002}
Abdalla, E.; Abellán, G.F.; Aboubrahim, A.; Agnello, A.; Akarsu, Ö; Akrami, Y.; ; Pettorino, V. Cosmology intertwined: A review of the particle physics, astrophysics, and cosmology associated with the cosmological tensions and anomalies
{\em J. High Energy Astrophys.} {\bf 2022}, {\em 34}, 49.

\bibitem[Aghanim, N. et al.(2020)]{10.1051/0004-6361/201833910}
Aghanim, N. Planck 2018 results. VI. Cosmological parameters. {\em Astron. Astrophys.} {\bf 2020}, {\em 641}, A6.

\bibitem[Di Valentino, E. and Mena, O. and Pan, S. and Visinelli, L. and Yang, W. and Melchiorri, A. and Mota, D. F. and Riess, A. G. and Silk, J.(2021)]{10.1088/1361-6382/ac086d}
Di Valentino, E.; Mena, O.; Pan, S.; Visinelli, L.; Yang, W.; Melchiorri, A.; Mota, D.F.; Riess, A.G.; Silk, J. In the realm of the Hubble tension—A review of solutions. {\em Class. Quantum Grav.} {\bf 2021}, {\em 38}, 153001.

\bibitem[Ivanov, M. M. and Ali-Haïmoud, Y. and Lesgourgues, J.(2020)]{10.1103/PhysRevD.102.063515}
Ivanov, M.M.; Ali-Haïmoud, Y.; Lesgourgues, J. $H_0$ tension or $T_0$ tension? {\em Phys. Rev. D.} {\bf 2020}, {\em 102}, 063515.

\bibitem[Pesce, D. W.and Braatz, J. A. and Reid, M. J. and Riess, A. G. Scolnic, D. and Condon, J. J. and Gao, F. and Henkel, C. and Impellizzeri, C. M. V. and Kuo, C. Y.(2020)]{10.3847/2041-8213/ab75f0}
Pesce, D.W.; Braatz, J.A.; Reid, M.J.; Riess, A.G.; Scolnic, D.; Condon, J.J.; Gao, F.; Henkel, C.; Impellizzeri, C.M.V.; Kuo, C.Y. The Megamaser Cosmology Project. XIII. Combined Hubble Constant Constraints. {\em  Astrophys. J. Lett.} {\bf 2020}, {\em 891}, L1.

\bibitem[Moresco, M. and Jimenez, R. and Verde, L. and Cimatti, A. and Pozzetti, L. and Maraston, C. and Thomas, D.(2016)]{10.1088/1475-7516/2016/12/039}
Moresco, M.; Jimenez, R.; Verde, L.; Cimatti, A.; Pozzetti, L.; Maraston, C.; Thomas, D. Constraining the time evolution of dark energy, curvature and neutrino properties with cosmic chronometers. {\em J. Cosmol. Astropart. Phys.} {\bf 2016}, {\em 12}, 039.

\bibitem[Foidl, H. and Rindler-Daller, T.(2024)]{10.1051/0004-6361/202348955}
Foidl, H.; Rindler-Daller, T. A proposal to improve the accuracy of cosmological observables and address the Hubble tension problem. {\em Astron. Astrophys.} {\bf 2024}, {\em 686}, A210.

\bibitem[Freedman, W. L.(2021)]{10.3847/1538-4357/ac0e95}
Freedman, W.L. Measurements of the Hubble Constant: Tensions in Perspective. {\em  Astrophys. J.} {\bf 2021}, {\em 919}, 16.

\bibitem[Freedman, W. L. and Madore, B. F. and Hatt, D. and Hoyt, T. J. and Jang, I. S. and Beaton, R. L. and Burns, C. R. and Lee, M. G. and Monson, A. J. and Neeley, J. R.(2019)]{10.3847/1538-4357/ab2f73}
Freedman, W.L.; Madore, B.F.; Hatt, D.; Hoyt, T.J.; Jang, I.S.; Beaton, R.L.; Burns, C.R.; Lee, M.G.; Monson, A.J.; Neeley, J.R. The Carnegie-Chicago Hubble Program. VIII. An Independent Determination of the Hubble Constant Based on the Tip of the Red Giant Branch. {\em Astrophys. J.} {\bf 2019}, {\em 882}, 34.

\bibitem[Jensen, J. B., Blakeslee, J. P., Cantiello, M., Cowles, M., Anand, G. S., Tully, R. B., and Raimondo, G. (2025).]
{10.3847/1538-4357/addfd6}
Jensen, J.B.; Blakeslee, J.P.; Cantiello, M.; Cowles, M.; An, G.S.; Tully, R.B.; Raimondo, G.  
The TRGB− SBF Project. III. Refining the HST Surface Brightness Fluctuation Distance Scale Calibration with JWST. 
{\em Astrophys. J.} {\bf 2019}, {\em 987}, 87.


\bibitem[Knox, L. and Millea, M.(2020)]{10.1103/PhysRevD.101.043533}
Knox, L.; Millea, M. Hubble constant hunter’s guide. {\em Phys. Rev. D} {\bf 2020}, {\em 101}, 043533.

\bibitem[Riess, A. G. and Anand, G. S. and Yuan, W. and Casertano, S. and Dolphin, A. and Macri, L. M. and Breuval, L. and Scolnic, D. and Perrin, M. and Anderson, R. I.(2024)]{10.3847/2041-8213/ad1ddd}
Riess, A.G.; Anand, G.S.; Yuan, W.; Casertano, S.; Dolphin, A.; Macri, L.M.;  Breuval, L.;  Scolnic, D.; Perrin, M.; Anderson, R.I. JWST Observations Reject Unrecognized Crowding of Cepheid Photometry as an Explanation for the Hubble Tension at 8$\sigma$ Confidence. {\em  Astrophys. J. Lett.} {\bf 2024}, {\em 962}, L17.

\bibitem[Bernal, J, L. and Verde, L, and Riess, A, G.(2016)]{10.1088/1475-7516/2016/10/019}
Bernal, J.L.; Verde, L.; Riess, A.G. The trouble with $H_0$. {\em J. Cosmol. Astropart. Phys.} {\bf 2016}, {\em 10}, 019.

\bibitem[Jia, J. and Niu, J. and Qiang, D. and Wei, H.(2025)]{https://arxiv.org/abs/2504.13380}
Jia, J.; Niu, J.; Qiang, D.; Wei, H. Alleviating the Hubble Tension with a Local Void and Transitions of the Absolute Magnitude. {\em Phys. Rev. D} {\bf 2025}, \textit{112}, 043507.

\bibitem[Poulin, V. and Smith, T. L. and Karwal, T. and Kamionkowski, M.(2019)]{10.1103/PhysRevLett.122.221301}
Poulin, V.; Smith, T.L.; Karwal, T.; Kamionkowski, M. Early Dark Energy can Resolve the Hubble Tension. {\em Phys. Rev. Lett.} {\bf 2019}, {\em 122}, 221301.

\bibitem[Desmond, H. and Jain, B. and Sakstein, J.(2019)]{10.1103/PhysRevD.100.043537}
Desmond, H.; Jain, B.; Sakstein, J. A local resolution of the Hubble tension: The impact of screened fifth forces on the cosmic distance ladder. {\em Phys. Rev. D} {\bf 2019}, {\em 100}, 043537.

\bibitem[Kreisch, C. D. and Cyr-Racine, F. and Doré, O.(2020)]{10.1103/PhysRevD.101.123505}
Kreisch, C.D.; Cyr-Racine, F.; Doré, O. Neutrino puzzle: Anomalies, interactions, and cosmological tensions. {\em Phys. Rev. D} {\bf 2020}, {\em 101}, 123505.

\bibitem[Sch{\"o}neberg, N. and Murgia, R. and Gariazzo, S. and Nesseris, S. and Nunes, R. C. and Renzi, A. and Vagnozzi, S. and Di Valentino, E.(2022)]{10.1103/PhysRevD.105.103511}
Sch{\"o}neberg, N.; Murgia, R.; Gariazzo, S.; Nesseris, S.; Nunes, R.C.; Renzi, A.; Vagnozzi, S.; Di Valentino, E. The Hubble tension: A global fit to cosmological data. {\em Phys. Rev. D} {\bf 2022}, {\em 105}, 103511.

\bibitem[Di Valentino, E., Said, J. L., Riess, A., Pollo, A., Poulin, V., Gómez-Valent, A., and Valls-Gabaud, D. (2025).]
{10.1016/j.dark.2025.101965}
Di Valentino, E.; Said, J.L.; Riess, A.; Pollo, A.; Poulin, V.; Gómez-Valent, A.; Valls-Gabaud, D.
The CosmoVerse White Paper: Addressing observational tensions in cosmology with systematics and fundamental physics. { \em Phys. Dark Universe} {\bf 2025}, {\em 49}, 101965.

\bibitem[Feeney, S. M. and Mortlock, D. J. and Dalmasso, N.(2018)]{10.1093/mnras/sty418}
Feeney, S.M.; Mortlock, D.J.; Dalmasso, N. Clarifying the Hubble constant tension with a Bayesian hierarchical model of the local distance ladder. {\em Mon. Not. R. Astron. Soc.} {\bf 2018}, {\em 476}, 3861--3882.

\bibitem[Lemos, P. and Lee, E. and Efstathiou, G. and Gratton, S.(2019)]{10.1093/mnras/sty3082}
Lemos, P.; Lee, E.; Efstathiou, G.; Gratton, S. Model independent $H(z)$ reconstruction using the cosmic inverse distance ladder. {\em Mon. Not. R. Astron. Soc.} {\bf 2019}, {\em 483}, 4803--4810.

\bibitem[Handley, W.(2021)]{10.1103/PhysRevD.103.L041301}
Handley, W. Curvature tension: evidence for a closed universe. {\em Phys. Rev. D} {\bf 2021}, {\em 103}, L041301.

\bibitem[Liu, G. and Wang, Y. and Zhao, W.(2024)]{10.1016/j.physletb.2024.138717}
Liu, G.; Wang, Y.; Zhao, W. Testing the consistency of early and late cosmological parameters with BAO and CMB data. {\em Phys. Lett. B} {\bf 2024}, {\em 854}, 138717.

\bibitem[Han, T. and Jin, S. and Zhang, J. and Zhang, X.(2024)]{10.1140/epjc/s10052-024-12999-w}
Han, T.; Jin, S.; Zhang, J.; Zhang, X. A comprehensive forecast for cosmological parameter estimation using joint observations of gravitational waves and short $\gamma$-ray bursts. {\em  Eur. Phys. J.} {\bf 2024}, {\em 84}, 663.

\end{thebibliography}


\isAPAandChicago{}{%

}

%


\PublishersNote{}
\end{adjustwidth}
\end{document}